Book Review of

Evolutionary and Interpretive Archaeologies

Edited by Ethan E. Cochrane and Andrew Gardner

Published by the University College London Institute of Archaeology in partnership with Left Coast Press (http://www.lcoastpress.com/book.php?id=325)

For *Kiva: Journal of Southwestern Anthropology and History*

Requested by Jeryll Moreno

Liane Gabora
University of British Columbia

and

Carl P. Lipo
California State University Long Beach

*Evolutionary and Interpretive Archaeologies*, edited by Ethan E. Cochrane and Andrew Gardner, grew out of a seminar at the Institute for Archaeology at University College London in 2007. It consists of 15 chapters by archaeologists who self-identify themselves as practitioners who emphasize the benefits of "evolutionary" or "interpretive" approaches to the study of the archaeological record. While the authors' theoretical views are dichotomous, the editors' aim for the book as a whole is not to expound on the differences between these two kinds of archaeology but to bring forward a richer understanding of the discipline and to highlight areas of mutual concern. Some chapters come across as a bit of a sales pitch, but the majority of the contributions emphasize how each approach can be productively used to address the goals of the other. The book seeks to contribute to a mutually beneficial and more productive discipline, and overall, it succeeds in this effort.

The first chapter, by the editors, discusses the history and present state of the divide between evolutionary and interpretive approaches, and outlines the stance they each take with respect to a number of unifying themes in archaeological research. The first theme focuses on the subject matter of archaeology. Evolutionary approaches, they claim, aim to characterize patterns of stability and change in the distribution of artifact variants using a conceptual framework based in Darwinian evolutionary theory. Interpretive approaches, in contrast, which emerged as a reaction against processualist reductionism, while widely variable in their details, tend to emphasize the role and perception of acting individuals. While interpretive accounts of the past view are commonly viewed as less scientific in comparison to evolutionary accounts, those who pursue interpretations forsake falsifiable explanations for reconstructions of the social, psychological and cultural contexts in which actors interacted to produce history. A second, related theme centers on a discussion of methods, though this topic is a minor one in the practice of interpretive archaeology. A third theme explores how evolutionary approaches tend to be generalizing whereas interpretive approaches tend to be particularizing. A final theme delves into the different views on the nature of existence and how these affect the characteristics of evolutionary and interpretive archaeology.

The remainder of the book following this introductory chapter is broken into three Parts. Part One, titled 'Theoretical Concerns', consists of four chapters that aim to characterize the theoretical groundwork underlying evolutionary and interpretive approaches. The first, by Cochrane, explains how methods such as seriation and engineering analysis are used to arrange and describe artifacts using concepts generalized from evolutionary theory and memetics. Here, the effort to bridge the conceptual paradigms falters somewhat, since memetics is a problematic derivation of evolutionary theory that has largely failed to lead to significantly new understandings of cultural change. A subsequent chapter by Gardner argues that archaeology is incomplete without taking into account the meaning of material culture as it is actively produced by those who make, use, and consume it. Gardner emphasizes that interpretive archaeologies use agency theory to understand the relationships between acting individuals, societies and institutions. The study of these contexts clearly has some relevancy in evolutionary based cultural transmission studies, since these interactions form the environment in which cultural inheritance takes place, and this must ultimately be considered in any dynamically sufficient explanatory framework. Bentley's chapter discusses the often-confused concepts of style and function, and argues how concepts from population genetics such as drift and selection can be applied to cultural variability. The last chapter of Part One by Sillar discusses creativity and agency with respect to the Inca state. Sillars argues that innovation does not occur *de novo,* and thus a full explanation of change must include an understanding the motivations and intensions unique to particular groups in particular times and places. While interesting, his explanation makes use of circular reasoning in which new variants are used to signal the intentions of those who create them.

Part Two, titled 'Contexts of Study', consists of seven chapters that address a variety of specific concepts that are used to make sense of the archaeological record. The chapters by James and Layton focus on notions of violence and conflict. Both chapters argue that archaeological explanations must link specific interpretations of the archaeological record using broadly conceived evolutionary frameworks that address perceived occurrences of violence while recognizing the role of cultural structures and biases. The chapter by Sommer discusses interpretation of the archaeological record in terms of cultural groups, tribes, and ethnicity. The next chapter by Glatz, Candler, and Steele discusses how distributions expected by neutral theory (as per population genetic explanations) can be used to interpret patterns of ceramic production at Bogazköy-Hattusa. A chapter by Whitehouse asks whether it is possible to reconcile biological and cultural approaches to embodiment, a topic that relies on principles argued to have general applicability. Tehrani's chapter argues that cladistics, a method borrowed from (and tailored to) biology for arranging sets of entities into branching trees of relatedness, has been applied to the study of historical relationships amongst artifacts. The concluding chapter in this section by Hamilton discusses issues pertaining to archaeological aspects of the complex ways in which humans understand and interact with physical landscapes, though it is unclear how this might accomplished in any specific empirical case.

Part Three, titled 'Future Directions', consists of three chapters that integrate concerns discussed earlier in the book and attempt to delineate the distinctive features of a coherent, if multi-stranded approach. Colleran and Mace argue that there has been a tendency to characterize evolutionary and interpretive approaches in extreme terms that do not do justice to the richness of these traditions. A chapter by Johnson discusses evolutionary archaeology from an interpretive perspective, and attempts to outline the goals and intellectual aims of archaeology as a discipline. In the final chapter, Shennan argues for a Darwinian evolution-based scientific archaeology, and speculates about the directions in which the field is headed.

Overall, the book is successful in its mission to highlight and integrate different approaches to the study of the archaeological record. As one reads through the chapters, it is clear that the two approaches to archaeological research are far from simplistic, and are not as dichotomous as the title suggests. This may produce a degree of confusion for anyone seeking a clear and useful answer to the question of which approach leads to superior interpretations. Interpretive concepts such as violence and relatedness are readily conceived in evolutionary terms, but it is less clear whether 'evolutionary' notions such as style and function can be profitably used in the construction of interpretations about personal motivation. It is also not clear that evolutionary approaches are better off with the introduction of ideas derived from interpretive efforts. One can argue that increasing the number of parts used in any conceptual frameworks can lead to "better" theory since the world is always more complex than the models we use to explain it. However, this approach to theory construction confuses the real world with the tools we use to explain it: the tools are of our construction that are merely sufficient to account for the world to the degree to which we demand it to be explained. Thus, arbitrarily throwing in more complexity does not necessarily lead to increased explanatory power. For example, evolutionary accounts of populations interacting on landscapes may or may not benefit from the 'hypercomplexity' involved in emic understandings of landscapes.

In this sense, the book is not entirely successful. Its failure has much to do with the way "evolutionary" and "interpretative" are treated as empirical entities rather than labels for analytic classes. Since labels are the product of classification, the classificatory units depend entirely on what one is trying to do. Lacking such a purpose, units are meaningless. Instead of starting with evolutionary and interpretive archaeology as a given, the editors could have identified the contributions with respect to specific goals in the explanation of the record. The discussion then might have distinguished those projects in which science is viewed as merely rigorous study, versus those that view it as the development and testing of empirically falsifiable theories. We suspect this might reveal that it is not necessarily the case that all parts of a single argument by a single author can readily meet the definition of one kind of archaeology or the other. There might still be two categories that include interpretation (defined as systematic descriptions of the causal agents that are recognized/asserted as playing a role in observed empirical phenomena) but this class of approach would be functionally distinguished from scientific archaeology (those studies that seek to use theory to generate observations that can be falsified). What one would find is that most of the chapters in this volume that are more or less "evolutionary" do not necessarily end up in the class of scientific efforts. For example, while making use of evolutionary terms, Bogazkoy-Hattusa's chapter on neutral traits in pottery types remains unconcerned about the linkage between the concepts used, measurement units (here, ceramic taxonomy), and the empirical record, and thus does not generate potentially falsifiable statements. Herein lies the source of the confusing overlap between what are initially described as distinct approaches to archaeology. Lacking attention to linking theory with empirical expectations of the archaeological record, the volume's evolutionary archaeology is structurally identical to interpretive accounts. Absent from the volume are concerns over measurement: how do we know we are measuring the world in a way that we can make statements that can be wrong? For the most part, the chapters assume the existence of 'data' in the archaeological record with standard practices and units serving as the means of investigation. However, when falsifiability is of little concern, one narrative is as good as another despite the language used in their construction. In these cases, evolutionary theory is simply used as an interpretive algorithm. Appearing to be evolutionary is not the same as producing a scientific explanation. In the end, the *practical* differences of the 'interpretive' and 'evolutionary' approaches to archaeology in this volume are much smaller than the ontological divide that separates them.

The book does not address serious concerns that have been raised concerning the application of biological origins of evolutionary theory to the social sciences (Atran, 2001; Fracchia & Lewontin, 1999; Gabora, 2004, 2006, 2008, 2011; Kauffman, 1999; Mayr, 1996; Tëmkin & Eldredge, 2007). In LG's view, some of the chapters display evidence of misunderstanding and mis-application of evolutionary theory, and a tendency to equate evolution with Darwinism, though there exist non-Darwinian evolutionary approaches to archaeology (e.g., Lipo, 2006; Gabora *et al*., 2011; Veloz, Tëmkin & Gabora, 2012). Such approaches capitalize on scientific tools and techniques to incorporate the kinds of contextual, nondeterministic factors as well as motivational factors and cognitive and processes by which humans extract meaning and forge understanding, which as interpretive archaeologists argue profoundly impact patterns of stability and change in the design of human-made objects.

Despite this deficiency, the volume demonstrates how evolutionary language can be applied to a wide variety of phenomena traditionally considered in the domain of anthropological narratives of agents and their motivations. In and of itself, this is a valuable contribution to the discipline. Using evolutionary theory does not limit one to esoteric details that lay outside the interests of most anthropologists. A variety of topics ranging from violence to the body to ethnicity to creativity can be cast in terms of evolutionary concepts. A major challenge is to characterize artifacts and their changes over time in terms of meaningful measurement units, incorporating not just the sorts of vertically transmitted perceptual attributes that are amenable to a biological inheritance model, but also conceptual factors in adaptive cultural change such as analogical transfer, competition, or complementarity amongst artifacts.

On the whole, the book is interesting, informative, and definitely worth reading. It demonstrates the variety of complementary approaches by which, using the remnants of our ancestors' efforts to master their world, express themselves, and survive, we are piecing together a coherent picture of how our humanness emerged. If combined with attention to measurement, and more formal (as opposed to pop science) understanding of the strengths and weaknesses of how evolutionary theory is currently being applied to culture, it may be a pivotal stepping-stone toward the archaeology of the future.

**Acknowledgements**

This work was funded by grants to Gabora from the National Science and Engineering Research Council of Canada, and the Fund for Scientific Research of Flanders, Belgium. Lipo would like to thank Mark E. Madsen and Timothy D. Hunt for fruitful discussions in this area of archaeological inquiry.